\documentclass[prd,aps,preprintnumbers,showpacs,amsfonts,floatfix,nofootinbib,superscriptaddress]{revtex4}
\usepackage{latexsym,graphicx,epsfig,psfrag,here}
\begin{document}
\preprint{DSF-2005/1 (Napoli)}

\title{Semileptonic decays of $B_c$ mesons
into charmonium states in a relativistic quark model}

\author{Mikhail A. Ivanov}
\affiliation{Bogoliubov Laboratory of Theoretical Physics, \\
Joint Institute for Nuclear Research, 141980 Dubna, Russia}
\author{Juergen G. K\"{o}rner}
\affiliation{Institut f\"{u}r Physik, Johannes Gutenberg-Universit\"{a}t, \\
D-55099 Mainz, Germany}
\author{Pietro Santorelli}
\affiliation{Dipartimento di Scienze Fisiche, Universit{\`a} di
Napoli "Federico II", Italy\\
Istituto Nazionale di Fisica Nucleare, Sezione di Napoli, Italy}

\vspace{0.4cm}

\begin{abstract}
We use the framework of a relativistic constituent quark model
to study the semileptonic transitions of the $B_c$ meson into
$(\bar c c)$ charmonium
states where $(\bar c c)=\eta_c\,(^1S_0),$\,\, $J/\psi\, (^3S_1),$\,\,
$\chi_{c0}\, (^3P_0),$\,\,$\chi_{c1}\, (^3P_1),$\,\, $h_c\, (^1P_1),$\,\,
$\chi_{c2}\, (^3P_2),$\,\, $\psi\, (^3D_2)$.
We compute the $q^2$--dependence of all relevant form factors
and give predictions for their semileptonic $B_c$ decay modes including
also their $\tau$-modes. We derive a formula for the polar angle
distribution of the charged lepton in the $(l\nu_l)$ c.m. frame and compute
the partial helicity rates that multiply the angular factors in the
decay distribution. For the discovery channel
$B_c\to J/\psi(\rightarrow \mu^+ \mu^-) l \nu$ we compute the
transverse/longitudinal composition of the $ J/\psi$
which can be determined by an angular analysis of the decay
$ J/\psi \rightarrow \mu^+ \mu^-$. We compare our results with the
results of other calculations.
\end{abstract}

\maketitle

\section{Introduction}
\label{s:intro}

In 1998 the CDF Collaboration reported on the observation of the
bottom-charm $B_c$ meson at Fermilab
\cite{CDF}. The $B_c$ mesons were found in an analysis of the
semileptonic decays $B_c\to J/\psi l \nu$ with the $J/\psi$ decaying
into muon pairs. Values for
the mass and the lifetime of the $B_c$ meson were given as
$M(B_c)=6.40\pm 0.39\pm 0.13$ GeV and
 $\tau(B_c)=0.46^{+0.18}_{-0.16}({\rm stat})\pm 0.03({\rm syst})\cdot
10^{-12}$ s, respectively. First $B_c$ mesons are now starting to be seen also
in the Run II data from the Tevatron \cite{lucchesi04,cdf04}.
Much larger samples of $B_c$ mesons and
more information on their decay properties are expected from the current
Run II at the Tevatron and future experiments at the LHC starting in 2007.
In particular this holds true for the dedicated detectors BTeV and
LHCB which are especially designed for the analysis of B physics where one
expects to see up to $10^{10}$ $B_c$ events per year.

The study of the $B_c$ meson is of great interest due to some of
its outstanding features. It is the lowest bound state of two
heavy quarks (charm and bottom) with open (explicit) flavor. As far
as the bound state characteristics are concerned the $B_c$ meson
is quite similar to the $J^{PC}=0^{-+}$ states $\eta_c$ and $\eta_b$
in the charmonium ($c\bar c$-bound state) and
the bottomium ($b\bar b$-bound state) sector. However, the $\eta_c$
and $\eta_b$ have hidden (implicit) flavor and decay strongly and
electromagnetically whereas the $B_c$-meson decays weakly since it
lies below the $B\bar D$-threshold.

The $B_c$ meson and its decays have been widely studied
in the literature.
The theoretical status of the $B_c$-meson was reviewed in
\cite{Gershtein:1998mb}. The $B_c$ lifetime and decays were studied in
the pioneering paper \cite{Lusignoli:1990ky}. The exclusive semileptonic
and nonleptonic (assuming factorization) decays of the $B_c$-meson were calculated
in a potential model approach \cite{Chang:1992pt}. The binding energy and the wave function of
the $B_c$-meson were computed by using a flavor-independent
potential with the parameters fixed by the $c\bar c$ and $b \bar b$
spectra and decays. The same processes were also studied in the
framework of the Bethe-Salpeter equation in \cite{AMV}, and, in the
relativistic constituent quark model formulated on the light-front
in \cite{AKNT}. Three-point sum rules of QCD and NRQCD were analyzed
in \cite{KLO} and \cite{Kiselev:2000pp}
to obtain the form factors of the semileptonic decays
of $B^+_c\to J/\psi(\eta_c)l^+\nu$ and $B^+_c\to B_s(B_s^\ast)l^+\nu$.
As shown by the authors of \cite{Jenkins}, the form factors
parameterizing the $B_c$ semileptonic matrix elements can be
related to a smaller set of form factors if one exploits the decoupling of
the spin of the heavy quarks in the $B_c$ and in the mesons produced in the
semileptonic decays. The reduced form factors can be
evaluated as an overlap integral of the meson wave-functions which can be
obtained, for example, using a relativistic potential model. This
was done in \cite{Colangelo}, where the $B_c$ semileptonic form
factors were computed and predictions for semileptonic and
non-leptonic decay modes were given.

In \cite{Ivanov:2000aj} we focused on its
exclusive leptonic and semileptonic decays which are sensitive to
the description of long distance effects. From the semileptonic decays
one can obtain results on the corresponding two-body non-leptonic decay
processes in the so-called factorization approximation.
The calculations have been done within
our relativistic constituent quark model based on an effective
Lagrangian describing the
coupling of hadrons $H$ to their constituent quarks. The relevant
coupling strength is determined by the compositeness condition $Z_H=0$
\cite{SWH,EI} where $Z_H$ is the wave function renormalization
constant of the hadron $H$.

The relativistic constituent quark model was also employed in a
calculation of the exclusive rare decays
$B_c\to D(D^\ast)\bar l l$ \cite{Faessler:2002ut} and of the nonleptonic
decays $B_c\to D_s \overline {D^0}$ and $B_c\to D_s D^0$ \cite{Ivanov:2002un}.
In the latter case we confirmed that
the nonleptonic decays $B_c\to D_s \overline {D^0}$ and
$B_c\to D_s D^0$ are well suited to extract the CKM
angle $\gamma$ through amplitude relations, as was originally proposed in
\cite{masetti1992,fleischer2000}. The reason is that the branching fractions
into the two channels are of the same order of magnitude.

In this paper we continue the study of $B_c$ decay properties
and calculate the branching rates of the semileptonic decays
$B_c\to (\bar c c)\,l\nu$
with $(\bar c c)=\eta_c\,(^1S_0),$\,\, $J/\psi\, (^3S_1),$\,\,
$\chi_{c0}\, (^3P_0),$\,\,$\chi_{c1}\, (^3P_1),$\,\, $h_c\, (^1P_1),$\,\,
$\chi_{c2}\, (^3P_2),$\,\, $\psi\, (^3D_2)$. We compare our results
with the results of \cite{Chang:1992pt,Chang:2001pm} where
it was shown that these decay rates are quite sizable
and may be accessible in RUN II of Tevatron and/or the LHC.
Two-particle decays of the $B_c$-meson into charmonium
states have been studied before in \cite{Kiselev:2001zb} by using
the factorization of hard and soft contributions.
The weak decays of the $B_c$-meson to charmonium have been studied
in the framework of the relativistic quark model
based on the quasipotential approach in \cite{Ebert:2003cn}.
In this paper we compute all form factors of the above semileptonic
$B_c$-transitions
and give predictions for various semileptonic $B_c$ decay modes including
their $\tau$-modes.
From a general point of view we would like to remark that the semileptonic
decays of the $\tau$-lepton have been studied within perturbative
QCD. It has allowed one to determine the strong coupling
constant with a high accuracy (see  e.g. \cite{Korner:2000xk}).
We have improved on our previous calculation \cite{Ivanov:2000aj}
in that we no longer employ the so-called impulse approximation.
In the impulse approximation one assumes that the vertex functions depend
only on the loop momentum flowing through the vertex. Dropping the
impulse approximation means that the vertex function can also depend
on outer momenta according to the flow of momentum through the vertex.
A comparison with the results for the decays into the para- and
ortho-charmonium states $(\bar c c)=\eta_c\,(^1S_0),$\,\, $J/\psi$
$(^3S_1)$ \cite{Ivanov:2000aj}, which was done
in the impulse approximation, shows a $\approx 10\%$ downward effect
in the rates when the impulse approximation is dropped.

%%%%%%%%%%%%%%%%%%%%%%%%%%%%%%%%%%%%%%%%%%%%%%%%%%%%%%%%%%%%%%%%%%%%%%%%
%%%%%%%%%%%%%%%%%%%%%%%%%%%%%%%%%%%%%%%%%%%%%%%%%%%%%%%%%%%%%%%%%%%%%%%%

\section{Bound state representation of the charmonium states}
\label{s:bound}

The charmonium states treated in this paper
are listed in Table~\ref{tab:states}. We have also included the purported
$D$--wave state $\psi(3836)$ whose quantum numbers have not been
established yet. Table~\ref{tab:states} also
contains the quark currents used to describe the coupling of the respective
charmonium states to the charm quarks. The masses of the charmonium
states listed in Table~\ref{tab:states} are taken from the
PDG \cite{Eidelman:2004wy}.

\begin{table}[ht]
\begin{center}
\caption{The charmonium states $^{2S+1}L_{\,J}$.
We use the notation $\stackrel{\leftrightarrow}{\partial}=
\stackrel{\rightarrow}{\partial}-\stackrel{\leftarrow}{\partial}$.}
\label{tab:states}
\def\arraystretch{1.5}
\begin{tabular}{|c|c|c|c|}
\hline
quantum number & name & quark current & mass (GeV) \\
\hline\hline
 $J^{PC}=0^{-+}$ (S=0, L=0) & $^1S_0=\eta_c$ & $\bar q\, i\gamma^5\, q $ &
2.980 \\
\hline
 $J^{PC}=1^{--}$ (S=1, L=0) & $^3S_1=J/\psi$ & $\bar q\,\gamma^\mu\, q $ &
3.097 \\
\hline
 $J^{PC}=0^{++}$ (S=1, L=1) & $^3P_0=\chi_{c0}$ & $\bar q\, q $ &
3.415 \\
\hline
 $J^{PC}=1^{++}$ (S=1, L=1) & $^3P_1=\chi_{c1}$ &
$\bar q\, \gamma^\mu\gamma^5\, q $ & 3.511 \\
\hline
$J^{PC}=1^{+-}$ (S=0, L=1) & $^1P_1=h_c(1P)$ &
 $\bar q\, \stackrel{\leftrightarrow}{\partial}^{\,\mu} \gamma^5\, q $
& 3.526 \\
\hline
 $J^{PC}=2^{++}$ (S=1, L=1) & $^3P_2=\chi_{c2}$ &
$(i/2)\,\bar q\, \left(\gamma^\mu \stackrel{\leftrightarrow}{\partial}^{\,\nu}
      +\gamma^\nu \stackrel{\leftrightarrow}{\partial}^{\,\mu}\right)\,q $
& 3.557 \\
\hline
 $J^{PC}=2^{--}$ (S=1, L=2) & $^3D_2=\psi(3836)$ &
$(i/2)\,
\bar q\left(\gamma^\mu\gamma^5 \stackrel{\leftrightarrow}{\partial}^{\,\nu}
 + \gamma^\nu \gamma^5 \stackrel{\leftrightarrow}{\partial}^{\,\mu} \right)q $
& 3.836 \\
\hline
\end{tabular}
\end{center}
\end{table}

Next we write down the Lagrangian describing the interaction of the
charmonium fields with the quark currents.
We give also the definition of the one-loop self-energy or mass insertions
(called mass functions in the following)
$\widetilde\Pi(p^2)$ of the relevant charmonium fields.

We can be quite brief in the presentation of the technical details of
our calculation since it is patterned after the calculation presented
in \cite{Ivanov:2000aj} which contains more calculational details.
We treat the different spin cases $(S=0,1,2)$ in turn.

\vspace{1cm}

\noindent
\underline{{\bf Spin S=0:}}

\begin{eqnarray}
\label{eq:s=0}
{\cal L}_{\rm \, S=0}(x) &=& \frac{1}{2}\,\phi(x)(\Box-m^2)\phi(x)
+g\,\phi(x)\,J_q(x), \hspace{1cm} \Box=-\partial^\alpha\partial_\alpha.
\\
&&\nonumber\\
\Pi(x-y) &=& i\,g^2\,\langle\, T\left\{J_q(x)\,J_q(y)\right\}\,\rangle_0,
\nonumber\\
\widetilde\Pi(p^2)&=&\int d^4x\, e^{-ipx}\,\Pi(x)\equiv \frac{3\,g^2}{4\pi^2}
\widetilde\Pi_0(p^2),\nonumber\\
&&\nonumber\\
Z &= & 1-\widetilde\Pi^{\,\prime}(m^2)=
            1-\frac{3\,g^2}{4\pi^2}\widetilde\Pi^{\,\prime}_0(m^2)=0,\nonumber\\
&&\nonumber\\
J_q(x)&=&\int\!\!\!\int\! dx_1dx_2\, F_{\rm cc}(x,x_1,x_2)\,
\bar q(x_1)\,\Gamma\, q(x_2)\,\qquad (\Gamma=i\,\gamma^5,I),\nonumber\\
F_{cc}(x,x_1,x_2)&=&\delta\left(x-\frac{x_1+x_2}{2}\right)
\Phi_{cc}\left((x_1-x_2)^2\right),\nonumber\\
\Phi_{cc}\left(x^2\right)&=&\int\!\frac{d^4q}{(2\,\pi)^4}\, e^{-iqx}\,
\widetilde\Phi_{cc}\left(-q^2\right).\nonumber
\end{eqnarray}

$\widetilde\Pi^{\,\prime}(m^2)$ is the derivative of the mass function
$\widetilde\Pi(p^2)$.

\vspace{1cm}

\noindent
\underline{{\bf Spin S=1:}}

\begin{eqnarray}
\label{eq:s=1}
{\cal L}_{\rm\, S=1}(x) &=&-\, \frac{1}{2}\,\phi_\mu(x)(\Box-m^2)\phi^\mu(x)
+g\,\phi_\mu(x)\,J^\mu_q(x),\\
\partial_\mu\phi^\mu(x) &=& 0
\qquad ({\rm leaving \, three \, independent \, components}),
\nonumber\\
&&\nonumber\\
\Pi^{\mu\nu}(x-y) &=& -\,i\,g^2\,
\langle\, T\left\{J^\mu_q(x)\,J^\nu_q(y)\right\}\,\rangle_0,\nonumber\\
\widetilde\Pi^{\mu\nu}(p)&=&\int d^4x\, e^{-ipx}\,\Pi^{\mu\nu}(x)
=g^{\mu\nu}\,\widetilde\Pi^{(1)}(p^2)+p^\mu p^\nu \widetilde\Pi^{(2)}(p^2),
\nonumber\\
\widetilde\Pi^{(1)}(p^2) &\equiv & \frac{3\,g^2}{4\pi^2}\,\widetilde\Pi_1(p^2),
\qquad
Z = 1-\frac{3\,g^2}{4\pi^2}\widetilde\Pi^{\,\prime}_1(m^2)=0,\nonumber\\
&&\nonumber\\
J^\mu_q(x)&=&\int\!\!\!\int\! dx_1dx_2\, F_{\rm cc}(x,x_1,x_2)\,
\bar q(x_1)\,\Gamma^\mu\, q(x_2),\,
\nonumber\\
\Gamma^\mu &=&\gamma^\mu,\,\gamma^\mu\gamma^5,\,
\stackrel{\leftrightarrow}{\partial}^{\,\mu}\gamma^5.
\nonumber
\end{eqnarray}

The spin 1 polarization vector $\epsilon^{(\lambda)}_\mu(p)$ satisfies
the constraints:

$$
\def\arraystretch{2.0}
\begin{array}{ll}
\displaystyle \epsilon^{(\lambda)}_\mu(p) \, p^\mu = 0 &
\hspace*{1cm} {\rm transversality},
\\
\displaystyle \sum\limits_{\lambda=0,\pm}\epsilon^{(\lambda)}_\mu(p)
\epsilon^{\dagger\,(\lambda)}_\nu (p)
=-g_{\mu\nu}+\frac{p_\mu\,p_\nu}{m^2} &
\hspace*{1cm} {\rm completeness},
\\
\displaystyle\epsilon^{\dagger\,(\lambda)}_\mu \epsilon^{(\lambda')\,\mu}
=-\delta_{\lambda \lambda'} &
\hspace*{1cm}  {\rm orthonormality}.
\end{array}
$$

\vspace{1cm}

\noindent
\underline{{\bf Spin S=2:}}

\begin{eqnarray}
\label{eq:s=2}
{\cal L}_{\rm\, S=2}(x) &=&
\frac{1}{2}\,\phi_{\mu\nu}(x)(\Box-m^2)\phi^{\mu\nu}(x)
+g\,\phi_{\mu\nu}(x)\,J_q^{\mu\nu}(x).\\
\phi^{\mu\nu}(x) &=&\phi^{\nu\mu}(x) , \quad
\partial_\mu\phi^{\mu\nu}(x) = 0,    \quad
\phi^{\mu}_{\mu}(x) = 0, \quad
({\rm leaving \,\, 5 \,\,independent\,\, components}),
\nonumber\\
&&\nonumber\\
\Pi^{\mu\nu,\alpha\beta}(x-y) &=&
i\,g^2\,<T\left\{ J^{\mu\nu}_q(x)\,J^{\alpha\beta}_q(y) \right\}>_0,
\nonumber\\
&&\nonumber\\
\widetilde\Pi^{\mu\nu,\alpha\beta}(p)&=&
\int d^4x\, e^{-ipx}\,\Pi^{\mu\nu,\alpha\beta}(x)=\frac{1}{2}
\left(g^{\mu\alpha}\,g^{\nu\beta}+g^{\mu\beta}\,g^{\nu\alpha}\right)\,
\widetilde\Pi^{(1)}(p^2)
\nonumber\\
&+& g^{\mu\nu}\,g^{\alpha\beta}\,\widetilde\Pi^{(2)}(p^2)
 +( g^{\mu\nu}\,p^\alpha p^\beta
   +g^{\mu\alpha}\,p^\nu p^\beta + g^{\mu\beta}\,p^\nu p^\alpha)\,
\widetilde\Pi^{(3)}(p^2)
+p^\mu p^\nu p^\alpha p^\beta\,\widetilde\Pi^{(4)}(p^2) ,
\nonumber\\
&&\nonumber\\
\widetilde\Pi^{(1)}(p^2) &\equiv & \frac{3\,g^2}{4\pi^2}\,\widetilde\Pi_2(p^2),
\qquad
Z = 1-\frac{3\,g^2}{4\pi^2}\widetilde\Pi^{\,\prime}_2(m^2)=0,\nonumber\\
&&\nonumber\\
J^{\mu\nu}_q(x)&=&\int\!\!\!\int\! dx_1dx_2\, F_{\rm cc}(x,x_1,x_2)\,
\bar q(x_1)\,\Gamma^{\mu\nu}\, q(x_2),
\nonumber\\
\Gamma^{\mu\nu} &=&\frac{i}{2}\,
\left(\gamma^\mu\stackrel{\leftrightarrow}{\partial}^{\,\nu}
+\gamma^\nu\stackrel{\leftrightarrow}{\partial}^{\,\mu}\right),\,\,\,\,
\frac{i}{2}\,
\left(\gamma^\mu\gamma^5\stackrel{\leftrightarrow}{\partial}^{\,\nu}
     +\gamma^\nu\gamma^5\stackrel{\leftrightarrow}{\partial}^{\,\mu}\right).
\nonumber
\end{eqnarray}
The spin 2 polarization vector $\epsilon^{(\lambda)}_{\mu\nu}(p)$ satisfies
the constraints:
$$
\def\arraystretch{1.7}
\begin{array}{ll}
\displaystyle\epsilon^{(\lambda)}_{\mu\nu}(p) = \epsilon^{(\lambda)}_{\nu\mu}(p)&
\hspace*{1cm} {\rm  symmetry},
\\
\displaystyle\epsilon^{(\lambda)}_{\mu\nu}(p)\, p^\mu = 0 &
\hspace*{1cm} {\rm transversality},
\\
\displaystyle\epsilon^{(\lambda)}_{\mu\mu}(p) = 0 &
\hspace*{1cm} {\rm tracelessness},
\\
\displaystyle\sum\limits_{\lambda=0,\pm 1, \pm 2}\epsilon^{(\lambda)}_{\mu\nu}
\epsilon^{\dagger\,(\lambda)}_{\alpha\beta}
=\frac{1}{2}\left(S_{\mu\alpha}\,S_{\nu\beta}
                   +S_{\mu\beta}\,S_{\nu\alpha}\right)
  -\frac{1}{3}\,S_{\mu\nu}\,S_{\alpha\beta} &
\hspace*{1cm} {\rm completeness},
\\
\displaystyle\epsilon^{\dagger\,(\lambda)}_{\mu\nu} \epsilon^{(\lambda')\,\mu\nu}
=\delta_{\lambda \lambda'} &
\hspace*{1cm}  {\rm orthonormality},
\end{array}
$$
where
$$
S_{\mu\nu} = -g_{\mu\nu}+\frac{p_\mu\,p_\nu}{m^2}.
$$

We use the the local representation for
the quark propagator when calculating the Fourier-tansforms
of the mass functions. The local quark propagator is given by
\begin{equation}
\label{local}
S_q(x-y) = \langle\,T\{q(x)\,\bar q(y)\}\,\rangle_0=
\int\frac{d^4 k}{(2\,\pi)^4\,i}\,e^{-i k\cdot(x-y)}\,\widetilde S_q(k), \qquad
\widetilde S_q(k)=\frac{1}{m_q-\not\! k}.
\end{equation}

For the mass functions one needs to calculate the integrals
\begin{eqnarray*}
\widetilde\Pi_0(p^2) &=& -\,\int\frac{d^4k}{4\,\pi^2\,i}\,
\widetilde\Phi_{cc}^2(-k^2)\,
{\rm Tr}\left[\Gamma\,\widetilde S(k-p/2)\,\Gamma\,\widetilde S(k+p/2)\right],
\\
\Gamma(P,S) &=& i\,\gamma^5,\,\,I.
\\
&&\\
\widetilde\Pi^{\mu\nu}_1(p) &=& \int\frac{d^4k}{4\,\pi^2\,i}\,
\widetilde\Phi_{cc}^2(-k^2)\,
{\rm Tr}\left[\Gamma^\mu\,\widetilde S(k-p/2)\,\Gamma^\nu\,\widetilde S(k+p/2)\right],
\\
\Gamma^\mu(V,A,PV)
&=& \gamma^\mu,\,\,\gamma^\mu\gamma^5,\,\,2\,i\,k^\mu\gamma^5.
\\
&&\\
\widetilde\Pi^{\mu\nu,\alpha\beta}_2(p) &=& \int\frac{d^4k}{4\,\pi^2\,i}\,
\widetilde\Phi_{cc}^2(-k^2)\,
{\rm Tr}\left[\Gamma^{\mu\nu}\,\widetilde S(k-p/2)\,
              \Gamma^{\alpha\beta}\,\widetilde S(k+p/2)\right],
\\
\Gamma^{\mu\nu}(T,PT) &=& i\,\left(\gamma^\mu\,k^\nu +\gamma^\nu\,k^\mu\right),
\,\,i\,\left(\gamma^\mu\gamma^5\,k^\nu +\gamma^\nu\gamma^5\,k^\mu\right).
\end{eqnarray*}

The functional form of the vertex function $\widetilde\Phi_{cc}(-k^2)$
and the quark propagators $\widetilde S_q(k)$ can in
principle be determined from an analysis of the Bethe-Salpeter and
Dyson-Schwinger equations as was done e.g. in \cite{Ivanov:1998ms}.
In this paper, however, we choose a phenomenological approach
where the vertex functions are modelled by a Gaussian form, the size
parameters of which are determined by a fit to the leptonic and radiative
decays of the lowest lying charm and bottom mesons. For the quark propagators
we use the above local representation Eq.~(\ref{local}).

We represent the vertex function by
\begin{eqnarray*}
\widetilde\Phi_{cc}(-k^2) &=& e^{s_{cc}\,k^2}, \qquad
s_{cc}=\frac{1}{\Lambda_{cc}^2},
\end{eqnarray*}
where $\Lambda_{cc}$ parametrizes the size of the charmonium state.
The quark propagator can easily be calculated using Feynman
parametrization. One has
\begin{eqnarray*}
\widetilde S_q(k\pm p/2) &=& (m_q+ \not\! k~\pm \not\! p/2)
\int\limits_0^\infty\! d\alpha\, e^{-\alpha\,(m_q^2-(k-p/2)^2)}\, .
\end{eqnarray*}
We then transform to new $\alpha$-variables according to
\[
\alpha_i \to  2\,s_{cc}\,\alpha_i,
\]
and make use of the identity
\begin{eqnarray*}
\int\!\!\!\int\limits_0^\infty\! d\alpha_1d\alpha_2 \,f(\alpha_1,\alpha_2) &=&
\int\limits_0^\infty\! dt\,t\!
\int\!\!\!\int\limits_0^\infty\! d\alpha_1d\alpha_2 \,
\delta\left(1-\alpha_1-\alpha_2\right) \,f(t\alpha_1,t\alpha_2) \, .
\end{eqnarray*}
One then obtains
\begin{equation}
\label{bracket}
\widetilde\Pi(p) = \left \langle\,\,\frac{1}{c^2_t}\,\int\frac{d^4 k}{\pi^2\,i}\,
e^{(k+c_p\,p)^2/c_t}\,\, \frac{1}{4}\,{\rm Tr}(\cdots)\,\right\rangle \, ,
\end{equation}
where
\[
c_t=\frac{1}{2(1+t)s_{cc}}, \qquad c_p=\frac{t}{1+t}\left (\alpha-\frac{1}{2}\right )
\]

The symbol $<...>$ stands for the two--fold integral
\begin{eqnarray*}
\langle\cdots\rangle &=&\int\limits_0^\infty dt\frac{t}{(1+t)^2}
\int\limits_0^1 d\alpha\,e^{-2\,s_{cc}\,z}\,(\cdots),\\
&&\\
z &=& t\,\left[m_c^2-\alpha(1-\alpha)\,p^2\right]
-\frac{t}{1+t}\,\left(\alpha-\frac{1}{2}\right)^2\,p^2\,.
\end{eqnarray*}

It is then convenient to shift the loop momentum according to
$k\to k-c_p\,p$. The
ensuing tensor integrals can be expressed by scalar integrals according to
\begin{eqnarray*}
\int\frac{d^4k}{\pi^2\,i}\,f(-k^2)\,k^\mu k^\nu k^\alpha k^\beta &=&
\frac{1}{24}\,\left(g^{\mu\nu}g^{\alpha\beta}+g^{\mu\alpha}g^{\nu\beta}
+g^{\mu\beta}g^{\nu\alpha}\right)\int\frac{d^4k}{\pi^2\,i}\,f(-k^2)\,k^4\,,
\\
\int\frac{d^4k}{\pi^2\,i}\,f(-k^2)\,k^\mu k^\nu  &=&
\frac{1}{4}\,g^{\mu\nu}\int\frac{d^4k}{\pi^2\,i}\,f(-k^2)\,k^2\, .
\end{eqnarray*}

The remaining scalar integrals can be integrated to give
\begin{equation}
\frac{1}{c^2_t}\,\int\frac{d^4 k}{\pi^2\,i}\,e^{k^2/c_t}\,k^{2n}
=(-)^n\,(n+1)!\,c_t^n \, .
\end{equation}

For the mass functions one finally obtains

$$
\begin{array}{lll}
\bar q\,i\,\gamma^5\,q & : &
\widetilde\Pi(p^2)_P=\langle\, 2\,c_t+m^2_c+(1/4-c^2_p)\,p^2\, \rangle \\
\bar q\,q & : &
\widetilde\Pi(p^2)_S=\langle\, 2\,c_t-m^2_c+(1/4-c^2_p)\,p^2\,\rangle \\
\bar q\,\gamma^\mu\,q & : &
\widetilde\Pi(p^2)_V=\langle\, c_t+m^2_c+(1/4-c^2_p)\,p^2\, \rangle \\
\bar q\,\gamma^\mu\gamma^5\,q & : &
\widetilde\Pi(p^2)_A=\langle\, c_t-m^2_c+(1/4-c^2_p)\,p^2\, \rangle \\
\bar q\,\stackrel{\leftrightarrow}{\partial}^{\,\mu}\gamma^5\,q & : &
\widetilde\Pi(p^2)_{PV}=2\,c_t\,
\langle\, 3\,c_t+m^2_c+(1/4-c^2_p)\,p^2\, \rangle \\
\bar q\,
\frac{i}{2}\left(\stackrel{\leftrightarrow}{\partial}^{\,\mu}\gamma^\nu
+\stackrel{\leftrightarrow}{\partial}^{\,\nu}\gamma^\mu\right)\,q & : &
\widetilde\Pi(p^2)_{T}=2\,c_t\,
\langle\, c_t+m^2_c+(1/4-c^2_p)\,p^2\, \rangle \\
\bar q\,
\frac{i}{2}
\left(\stackrel{\leftrightarrow}{\partial}^{\,\mu}\gamma^\nu\gamma^5
 +\stackrel{\leftrightarrow}{\partial}^{\,\nu}\gamma^\mu\gamma^5\right)\,q &
: &
\widetilde\Pi(p^2)_{PT}=2\,c_t\,
\langle \,c_t-m^2_c+(1/4-c^2_p)\,p^2\, \rangle \\
\end{array}
$$
The mass functions $\widetilde\Pi_I(p^2)$ enter the compositeness condition
in the derivative form
$$
Z_I=1-\frac{3\,g^2_I}{4\,\pi^2}\,\widetilde \Pi_I^\prime(p^2) \, ,
$$
where the prime denotes differentiation with respect to $p^2$.
The differentiation of the mass functions result in
\begin{eqnarray*}
\widetilde\Pi^\prime(p^2)_P &=&
\langle\, -2\,s_{cc}\,\hat z\,[\,2\,c_t+m^2_c+(1/4-c^2_p)\,p^2\,]
+1/4-c^2_p\,\,\rangle \\
\widetilde\Pi^\prime(p^2)_S &=&
\langle -2\,s_{cc}\,\hat z\,[\,2\,c_t-m^2_c+(1/4-c^2_p)\,p^2\,]
+1/4-c^2_p\,\,\rangle \\
\widetilde\Pi^\prime(p^2)_V &=&
\langle\,  -2\,s_{cc}\,\hat z\,[\,c_t+m^2_c+(1/4-c^2_p)\,p^2\,]
+1/4-c^2_p\,\,\rangle \\
\widetilde\Pi^\prime(p^2)_A &=&
\langle\, -2\,s_{cc}\,\hat z\,[\,c_t-m^2_c+(1/4-c^2_p)\,p^2\,]
+1/4-c^2_p\,\,\rangle \\
\widetilde\Pi^\prime(p^2)_{PV} &=& 2\,c_t\,
\langle -2\,s_{cc}\,\hat z\,[\,3\,c_t+m^2_c+(1/4-c^2_p)\,p^2\,]
+ 1/4-c^2_p \,\,\rangle \\
\widetilde\Pi^\prime(p^2)_{T} &=& 2\,c_t\,
\langle\, -2\,s_{cc}\,\hat z\,[\,c_t+m^2_c+(1/4-c^2_p)\,p^2\,]
+1/4-c^2_p\,\, \rangle \\
\widetilde\Pi^\prime(p^2)_{PT} &=& 2\,c_t\,
\langle\, -2\,s_{cc}\,\hat z\,[\,c_t-m^2_c+(1/4-c^2_p)\,p^2\,]
 +1/4-c^2_p\,\,\rangle
\end{eqnarray*}
where
$$
\hat z=-t\,\alpha(1-\alpha)-\frac{t}{1+t}\,\left(\alpha-\frac{1}{2}\right)^2.
$$

%%%%%%%%%%%%%%%%%%%%%%%%%%%%%%%%%%%%%%%%%%%%%%%%%%%%%%%%%%%%%%%%%%%%%%%%%%
%%%%%%%%%%%%%%%%%%%%%%%%%%%%%%%%%%%%%%%%%%%%%%%%%%%%%%%%%%%%%%%%%%%%%%%%%%

\section{The semileptonic decays \boldmath{$B_c\to (\bar cc)\,+ l + \bar\nu$}}
\label{s:semilep}

Let us first write down the interaction Lagrangian which we need for the
calculation
of the matrix elements of the semileptonic decays
$B_c\to (\bar cc)+l + \bar\nu$. One has
\begin{eqnarray*}
{\cal L}_{\rm int}(x) &=&
g_{bc}\,B_c^-(x)\cdot J^+_{bc}(x)+g_{cc}\,\phi_{cc}(x)\cdot J_{cc}(x)
+\frac{G_F}{\sqrt{2}}\,V_{bc}\,(\bar c\,O^\mu\, b)\cdot (\bar l\, O_\mu\,\nu),
\\
&&\nonumber\\
J^+_{bc}(x)&=&\int\!\!\!\int\! dx_1dx_2\, F_{\rm bc}(x,x_1,x_2)\cdot
\bar b(x_1)\,i\,\gamma^5\, c(x_2),
\\
J_{cc}(x)&=&\int\!\!\!\int\! dx_1dx_2\, F_{\rm cc}(x,x_1,x_2)\cdot
\bar c(x_1)\,\Gamma_{cc}\, c(x_2),
\\
&&\\
F_{bc}(x,x_1,x_2)&=&\delta(x-c_1\,x_1-c_2\,x_2)
\,\Phi_{bc}\left((x_1-x_2)^2\right),\\
F_{cc}(x,x_1,x_2)&=&\delta\left(x-\frac{x_1+x_2}{2}\right)
\,\Phi_{cc}\left((x_1-x_2)^2\right),\\
&&\\
\Phi\left(x^2\right)&=&\int\!\frac{d^4q}{(2\,\pi)^4} e^{-iqx}
\widetilde\Phi\left(-q^2\right).
\end{eqnarray*}
Here we adopt the notation: $l=e^-,\,\mu^-,\,\tau^-$,
$\bar l=e^+,\,\mu^+,\,\tau^+$, $O^\mu=\gamma^\mu\,(1-\gamma^5)$,
$c_1=m_b/(m_b+m_c),\,c_2=m_c/(m_b+m_c)$.

The S-matrix element describing the semileptonic decays
$B_c\to (\bar cc)\,+l\bar\nu$ is written as
\begin{eqnarray*}
S_{B_c\to (\bar cc)} &=& i^3\,g_{bc}\,g_{cc}\,\frac{G_F}{\sqrt{2}}\,V_{bc}\,
\int\!\!\!\int\!\!\!\int\! dxdx_1dx_2\,B_c^-(x)\,\delta(x-c_1\,x_1-c_2\,x_2)\,
\Phi_{bc}\left((x_1-x_2)^2\right)\\
&\times&
\int\!\!\!\int\!\!\!\int\! dydy_1dy_2\,\phi_{cc}(y)\,
\delta\left(y-\frac{y_1+y_2}{2}\right)\,
\Phi_{cc}\left((y_1-y_2)^2\right)
\int\!dz \left(\bar l\,O_\mu\, \nu\right)_z \\
&\times& \langle\, T\left\{\bar b(x_1)\,i\,\gamma^5\,c(x_2)\cdot
\bar c(y_1)\,\Gamma_{cc}\,c(y_2)\cdot
\bar c(z)\,O^\mu\,c(z)\right\}\,\rangle_0.
\end{eqnarray*}

The matrix element is calculated in the standard manner.
We have
\begin{eqnarray*}
T_{B_c\to (\bar cc)}(p_1,p_2,k_l,k_\nu) &=&
i\,(2\pi)^4\,\delta(p_1-p_2-k_l-k_\nu)\,
M_{B_c\to (\bar cc)}(p_1,p_2,k_l,k_\nu),\\
&&\\
M_{B_c\to (\bar cc)}(p_1,p_2,k_l,k_\nu) &=&
\frac{G_F}{\sqrt{2}}\,V_{bc}\,{\cal M}^\mu(p_1,p_2)\,
\bar u_l(k_l)\,O^\mu\,u_\nu(k_\nu),\\
&&\\
{\cal M}^\mu(p_1,p_2) &=& -\,\frac{3\,g_{bc}\,g_{cc}}{4\,\pi^2}\,
\int\!\frac{d^4k}{\pi^2\,i}\,\widetilde\Phi_{bc}\left(-(k+c_2\,p_1)^2\right)\,
\widetilde\Phi_{cc}\left(-(k+p_2/2)^2\right)\,\\
&\times& \frac{1}{4}\,{\rm Tr}\left[\,i\,\gamma^5\,\widetilde S_c(k)\,
\Gamma_{cc}\,\widetilde S_c(k+p_2)\,O^\mu\,\widetilde S_b(k+p_1)\,\right]
\end{eqnarray*}
where $p_1$ and $p_2$ are the $B_c$ and $(\bar cc)$ momenta, respectively.
The spin coupling structure of the $(\bar{c} c)$--states is given by
$$
\Gamma_{cc}=i\,\gamma^5,\,\, I,\,\,
\epsilon^{\dagger}_\nu\gamma^\nu,\,\,
\epsilon^{\dagger}_\nu\gamma^\nu\gamma^5,\,\,
-\,2\,i\,\epsilon^{\dagger\,\nu} k_\nu\gamma^5,\,\,
2\,\epsilon^{\dagger}_{\nu\alpha}\,k^\nu \gamma^\alpha,\,\,
2\,\epsilon^{\dagger}_{\nu\alpha}\,k^\nu \gamma^\alpha\gamma^5.
$$

The calculation of the transition matrix elements $M^\mu$ proceeds along
similar lines as in the case of the mass functions. For the scalar vertex
functions one has
\begin{eqnarray*}
\widetilde\Phi_{bc}(-(k+c_2\,p_1)^2) &=& e^{s_{bc}\,(k+c_2\,p_1)^2}, \qquad
s_{bc}=\frac{1}{\Lambda_{bc}^2},
\\
\widetilde\Phi_{cc}(-(k+p_2/2)^2) &=& e^{s_{cc}\,(k+p_2/2)^2}, \qquad
s_{cc}=\frac{1}{\Lambda_{cc}^2},
\\&&\\
\widetilde S_q(k+p) &=& (m_q+\not\! k + \not\! p)
\int\limits_0^\infty\! d\alpha\, e^{-\alpha\,(m_q^2-(k+p)^2)}.
\end{eqnarray*}

Again we shift the parameters $\alpha_i (i=1,2,3)$ according to
\begin{eqnarray*}
\alpha_i &\to & (s_{bc}+s_{cc})\,\alpha_i, \\
&&\\
\int\limits_0^\infty\! d^3\alpha \,
f(\alpha_1,\alpha_2,\alpha_3) &=&
\int\limits_0^\infty\! dt\,t^2
\int\limits_0^\infty\! d^3\alpha \,
\delta\left(1-\sum_{i=1}^3\alpha_i\right) \,
f(t\alpha_1,t\alpha_1,t\alpha_3)
\end{eqnarray*}
where $d^3\alpha=d\alpha_1 d\alpha_2 d\alpha_3$. One then obtains
\begin{eqnarray*}
{\cal M}^\mu(p_1,p_2) &=&
\left\langle\,\,\frac{1}{c^2_t}\,\int\frac{d^4 k}{\pi^2\,i}\,
e^{(k+c_{p_1}\,p_1+c_{p_2}\,p_2)^2/c_t}\,\,
\frac{1}{4}\,{\rm Tr}(\cdots)\,\right\rangle \, , \\
&&\\
c_t&=&\frac{1}{(s_{bc}+s_{cc})(1+t)}, \\
c_{p_1}&=&\frac{c_2\,w_{bc}+t\,\alpha_1}{1+t}, \qquad
c_{p_2} = \frac{w_{cc}/2+t\,\alpha_2}{1+t},\\
w_{bc} &=&\frac{s_{bc}}{s_{bc}+s_{cc}},  \qquad
w_{cc}  = \frac{s_{cc}}{s_{bc}+s_{cc}}.
\end{eqnarray*}
where the symbol $<...>$ is related to the corresponding symbol
$<...>$ defined in Sec.~\ref{s:bound} Eq.~(\ref{bracket}). In the present case
the symbol $<...>$ stands for the four-fold integral
\begin{eqnarray*}
\langle\cdots\rangle &=&(s_{bc}+s_{cc})\cdot
\int\limits_0^\infty\! dt\,\frac{t^2}{(1+t)^2}
\int\limits_0^1\, d^3\alpha\,
\delta\left(1-\sum_{i=1}^3\alpha_i\right) \,
e^{-\,(s_{bc}+s_{cc})\,z}\,(\cdots)\, ,
\end{eqnarray*}
where
\begin{eqnarray*}
z &=& t\,\left[(\alpha_2+\alpha_3)\,m_c^2+\alpha_1\,m_b^2
-\alpha_1\alpha_3\,p_1^2-\alpha_2\alpha_3\,p_2^2-\alpha_1\alpha_2\,q^2\right]
\\
&+&\frac{1}{1+t}\,\left\{\frac{}{}
p_1^2\,\left[\,t\,w_{bc}\,c_2\,(2\,\alpha_1+\alpha_2-c_2)
+t\,w_{cc}\,\alpha_1/2-t\,\alpha_1\,(\alpha_1+\alpha_2)
+,w_{bc}\,c_2\,(w_{cc}/2-c_2+w_{bc}\,c_2)\right]
\right.\\
&+&
p_2^2\,\left[\,t\,w_{bc}\,c_2\,\alpha_2
+t\,w_{cc}\,(\alpha_1/2+\alpha_2-1/4)\,-t\,\alpha_2\,(\alpha_1+\alpha_2)
+w_{cc}/4\,(2\,w_{bc}\,c_2-1+w_{cc}\,)\right] \\
&+&
\left.
\,q^2\,\left[\,t\,(-w_{bc}\,c_2\,\alpha_2-w_{cc}\,\alpha_1/2+\alpha_1\alpha_2)
-w_{bc}\,w_{cc}\,c_2/2\,\right]
\frac{}{}\right\}.
\end{eqnarray*}

One then shifts the loop momentum according to
$k\to k-c_{p_1}\,p_1-c_{p_2}\,p_2$.

Our final results are given in terms of a set of invariant form factors
defined by
\begin{eqnarray}
{\cal M}^\mu(\,B_c\to (\bar cc)_{\,S=0}\,) &=&
P^\mu\,F_+(q^2)+q^\mu\,F_-(q^2),\label{ff0}\\
&&\nonumber\\
{\cal M}^\mu(\,B_c\to (\bar cc)_{\,S=1} \,) &=&
\frac{1}{m_{B_c}+m_{cc}}\,\epsilon^\dagger_\nu\,
\left\{\,
-\,g^{\mu\nu}\,Pq\,A_0(q^2)+P^\mu\,P^\nu\,A_+(q^2)
+q^\mu\,P^\nu\,A_-(q^2)
\right.\nonumber\\
&&
\left.
+i\,\varepsilon^{\mu\nu\alpha\beta}\,P_\alpha\,q_\beta\,V(q^2)\right\},
\label{ff1}\\
&&\nonumber\\
{\cal M}^\mu(B_c\to (\bar cc)_{\,S=2} ) &=&
\epsilon^\dagger_{\nu\alpha}\,
\left\{\,
g^{\mu\alpha}\,P^\nu\,T_1(q^2)
+P^\nu\,P^\alpha\,\left[\,P^\mu\,T_2(q^2)+q^\mu\,T_3(q^2)\,\right]
\right.\nonumber\\
&&
\left.
+i\,\varepsilon^{\mu\nu\delta\beta}\,P^\alpha\,P_\delta\,q_\beta\,T_4(q^2)
\right\},\label{ff2}\\
&&\nonumber\\
P &=&p_1+p_2, \qquad q=p_1-p_2.
\nonumber
\end{eqnarray}
In our results we have dropped an overall phase factors which is irrelevant
for the calculation of the decay widths.

The calculation of traces and invariant
integrations is done with help of FORM \cite{Vermaseren:2000nd}.
For the values of the model parameters
(hadron sizes $\Lambda_H$ and constituent quark masses $m_q$)
we use the values of \cite{Ivanov:2003ge}. The numerical evaluation of
the form factors is done in FORTRAN.

%%%%%%%%%%%%%%%%%%%%%%%%%%%%%%%%%%%%%%%%%%%%%%%%%%%%%%%%%%%%%%%%%%%%%%%%%%
%%%%%%%%%%%%%%%%%%%%%%%%%%%%%%%%%%%%%%%%%%%%%%%%%%%%%%%%%%%%%%%%%%%%%%%%%%

\section{Angular decay distributions}
\label{s:angdistr}

Consider the semileptonic decays
$B^-_c(p_1)\to (\bar cc)(p_2)+l(k_l)+\bar\nu(k_\nu)$ and
$B^+_c(p_1)\to (\bar cc)(p_2)+\bar l(k_l)+\nu(k_\nu).$
Recalling the expression for the matrix elements, one can write
\begin{eqnarray*}
M_{B^-_c\to\bar cc}(p_1,p_2,k_l,k_\nu) &=&
\frac{G_F}{\sqrt{2}}\,V_{bc}\, {\cal M}_\mu(p_1,p_2)\,
\bar u^{\lambda}_l(\vec k_l)\,O^\mu\,v^{\lambda'}_\nu(\vec k_\nu),
%\hspace{1cm} {\rm lepton\,\, in\,\, the\,\, final\,\, state}
\\
&&\\
M_{B^+_c\to\bar cc}(p_1,p_2,k_l,k_\nu) &=&
\frac{G_F}{\sqrt{2}}\,V_{bc}\, {\cal M}_\mu(p_1,p_2)\,
\bar u^{\lambda}_\nu(\vec k_\nu)\,O^\mu\,v^{\lambda'}_l(\vec k_l),
\\
\end{eqnarray*}
where $p_1$ and $p_2$ are the $B_c$ and $(\bar cc)$ momenta, respectively.

The angular decay distribution reads
\begin{equation}
\label{width2}
\frac{d\Gamma}{dq^2\,d\cos\theta}=
\frac{G_F^2}{(2\pi)^3}|V_{bc}|^2\cdot
\frac{(q^2-\mu^2)\,|{\bf p_2}|}{8\,m_1^2\,q^2}\cdot
L^{\mu\nu}\,H_{\mu\nu}
\end{equation}
where $\mu$ is the lepton mass and
$|{\bf p_2}|=\lambda^{1/2}(m_1^2,m_2^2,q^2)/(2\,m_1)$ is the momentum
of the $(\bar cc)$-meson in the $B_c$-rest frame.

$L^{\mu\nu}$ is the lepton tensor given by
\begin{equation}
\label{lepton-tensor}
L^{\mu\nu}_\mp =
\frac{1}{8}\, {\rm Tr}\left(O^\mu\not\! k_\nu O^\nu\not\! k_l\right)
=
 k_l^\mu\,k_\nu^\nu+k_l^\nu\,k_\nu^\mu
-g^{\mu\nu}\,\frac{q^2-\mu^2}{2}
\pm i\,\varepsilon^{\mu\nu\alpha\beta}\,k_{l\,\alpha}k_{\nu\,\beta},
\end{equation}

The lepton tensors $L^{\mu\nu}_-$ and $L^{\mu\nu}_+$ refer to the
$(l\bar \nu_l)$ and $(\bar l \nu_l)$ cases. They
differ in the sign of the parity--odd $\varepsilon$-tensor contribution.
The hadron tensor
$H_{\mu\nu}={\cal M}_\mu(p_1,p_2)\, {\cal M}^\dagger_\nu(p_1,p_2)$
is given by the corresponding tensor products of the transition matrix
elements defined above.

It is convenient to perform the Lorentz contractions in Eq.~(\ref{width2})
with the help of helicity amplitudes as described in \cite{Korner:1989qb}
and \cite{Korner:1982bi,Korner:vg}.
First, we define an orthonormal and complete helicity basis
$\epsilon^\mu(m)$ with the three spin 1 components orthogonal to
the momentum transfer $q^\mu$, i.e. $\epsilon^\mu(m) q_\mu=0$ for $m=\pm,0$,
and the spin 0 (time)-component $m=t$ with
$\epsilon^\mu(t)= q^\mu/\sqrt{q^2}$.

The orthonormality and completeness properties read
\begin{eqnarray}
&&
\epsilon^\dagger_\mu(m)\epsilon^\mu(n)=g_{mn} \hspace{1cm}
(m,n=t,\pm,0),
\nonumber\\
&&\label{orthonorm}\\
&&
\epsilon_\mu(m)\epsilon^{\dagger}_{\nu}(n)g_{mn}=g_{\mu\nu}
\nonumber
\end{eqnarray}
with $g_{mn}={\rm diag}\,(\,+\,,\,\,-\,,\,\,-\,,\,\,-\,)$.
We include the time component polarization vector $\epsilon^\mu(t)$
in the set because we want to include lepton mass effects in the following.

Using the completeness property we rewrite the contraction
of the lepton and hadron tensors in Eq.~(\ref{width2}) according to
\begin{eqnarray}
L^{\mu\nu}H_{\mu\nu} &=&
L_{\mu'\nu'}g^{\mu'\mu}g^{\nu'\nu}H_{\mu\nu}
= L_{\mu'\nu'}\epsilon^{\mu'}(m)\epsilon^{\dagger\mu}(m')g_{mm'}
\epsilon^{\dagger \nu'}(n)\epsilon^{\nu}(n')g_{nn'}H_{\mu\nu}
\nonumber\\
&&\nonumber\\
&=& L(m,n)g_{mm'}g_{nn'}H(m',n')
\label{contraction}
\end{eqnarray}
where we have introduced the lepton and hadron tensors in the space
of the helicity components
\begin{eqnarray}
\label{hel_tensors}
L(m,n) &=& \epsilon^\mu(m)\epsilon^{\dagger \nu}(n)L_{\mu\nu},
\hspace{1cm}
H(m,n) = \epsilon^{\dagger \mu}(m)\epsilon^\nu(n)H_{\mu\nu}.
\end{eqnarray}
The point is that the two tensors can be evaluated in two different
Lorentz systems. The lepton tensors $L(m,n)$ will be evaluated
in the $\bar l \nu$ or $ l \bar \nu$--c.m. system whereas the hadron tensors
$H(m,n)$
will be evaluated in the $B_c$ rest system.

%%%%%%%%%%%%%%%%%%%%%%%%%%%%%%%%%%%%%%%%%%%%%%%%%%%%%%%%%%%%%%%%%%%%%%%%

\subsection{Hadron tensor}

In the $B_c$ rest frame one has
\begin{eqnarray}
p^\mu_1 &=& (\,m_1\,,\,\,0,\,\,0,\,\,0\,)\,,
\nonumber\\
p^\mu_2 &=& (\,E_2\,,\,\,0\,,\,\,0\,,\,\,-|{\bf p_2}|\,)\,,
\\
q^\mu   &=& (\,q_0\,,\,\,0\,,\,\,0\,,\,\,|{\bf p_2}|\,)\,,
\nonumber
\end{eqnarray}
where
\begin{eqnarray*}
&&
E_2 = \frac{m_1^2+m_2^2-q^2}{2\,m_1}, \hspace{1cm}
q_0=\frac{m_1^2-m_2^2+q^2}{2\,m_1},\\
&&
E_2+q_0=m_1, \hspace{1cm}
q_0^2=q^2+|{\bf p_2}|^2,\hspace{1cm}
|{\bf p_2}|^2+E_2\,q_0=\frac{1}{2}(m_1^2-m_2^2-q^2).
\end{eqnarray*}

In the $B_c$-rest frame the polarization vectors of the effective
current read
\begin{eqnarray}
\epsilon^\mu(t)&=&
\frac{1}{\sqrt{q^2}}(\,q_0\,,\,\,0\,,\,\,0\,,\,\,|{\bf p_2}|\,)
\,,\nonumber\\
\epsilon^\mu(\pm) &=&
\frac{1}{\sqrt{2}}(\,0\,,\,\,\mp 1\,,\,\,-i\,,\,\,0\,)\,,
\label{hel_basis}\\
\epsilon^\mu(0) &=&
\frac{1}{\sqrt{q^2}}(\,|{\bf p_2}|\,,\,\,0\,,\,\,0\,,\,\,q_0\,)\,.
\nonumber
\end{eqnarray}
Using this basis one can express the helicity components of the hadronic
tensors through the invariant form factors defined in
Eqs.~(\ref{ff0}-\ref{ff2}). We treat the three spin cases in turn.

\vspace{0.5cm}
\noindent
(a)\underline{ $B_c\to (\bar cc)_{\,S=0}$ transition}:

\begin{equation}
\label{helS0a}
H(m,n)= \left(\epsilon^{\dagger \mu}(m){\cal M}_\mu\right)\cdot
        \left(\epsilon^{\dagger \nu}(n){\cal M}_\nu\right)^\dagger\equiv
H_mH^\dagger_n
\end{equation}
The helicity form factors $H_m$ can be expressed in terms of the
invariant form factors. One has
\begin{eqnarray}
H_t &=& \frac{1}{\sqrt{q^2}}(Pq\, F_+ + q^2\, F_-)\,,
\nonumber\\
H_\pm &=& 0\,,
\label{helS0b}\\
H_0 &=& \frac{2\,m_1\,|{\bf p_2}|}{\sqrt{q^2}} \,F_+ \,.
\nonumber
\end{eqnarray}

\vspace{0.5cm}
\noindent
(b) \underline{$B_c\to (\bar cc)_{\,S=1}$ transition}: \\
\vspace{-0.1cm}

The nonvanishing helicity form factors are given by
\begin{equation}
H_m =\epsilon^{\dagger \mu}(m){\cal M}_{\mu\alpha}\epsilon_2^{\dagger\alpha}(m)
\quad {\rm for} \quad m=\pm,0
\end{equation}
and
\begin{equation}
H_t =\epsilon^{\dagger \mu}(t){\cal M}_{\mu\alpha}\epsilon_2^{\dagger\alpha}(0)
\end{equation}

\noindent As in Eq.~(\ref{helS0a}) the hadronic tensor is given by
$H(m,n)= H_mH^\dagger_n$.

In order to express the helicity form factors in terms of the invariant
form factors Eq.~(\ref{ff1}) one needs to specify the helicity components
$\epsilon_2(m)$ $(m=\pm,0)$ of the polarization vector of the
$(\bar cc)_{S=1}$ state. They are given by
\begin{eqnarray}
\epsilon^\mu_2(\pm) &=&
\frac{1}{\sqrt{2}}(0\,,\,\,\pm 1\,,\,\,-i\,,\,\,0\,)\,,
\nonumber\\
&&\label{pol_S1}\\
\epsilon^\mu_2(0) &=&
\frac{1}{m_2}(|{\bf p_2}|\,,\,\,0\,,\,\,0\,,\,\,-E_2\,)\,.
\nonumber
\end{eqnarray}
They satisfy the orthonormality and completeness conditions:
\begin{eqnarray}
&&
\epsilon_2^{\dagger\mu}(r)\,\epsilon_{2\mu}(s)=-\delta_{rs},
\nonumber\\
&&\\
&&
\epsilon_{2\mu}(r)\,\epsilon^\dagger_{2\nu}(s)\,\delta_{rs}=
-g_{\mu\nu}+\frac{p_{2\mu}p_{2\nu}}{m_2^2}.\nonumber
\end{eqnarray}

The desired relations between the helicity form factors and the
invariant form factors are then

\begin{eqnarray}
H_t &=&
\epsilon^{\dagger \mu}(t)\,\epsilon_2^{\dagger \alpha}(0)
\,{\cal M}_{\mu\alpha}
\,=\,
\frac{1}{m_1+m_2}\frac{m_1\,|{\bf p_2}|}{m_2\sqrt{q^2}}
\left(P\cdot q\,(-A_0+A_+)+q^2 A_-\right),
\nonumber\\
H_\pm &=&
\epsilon^{\dagger \mu}(\pm)\,\epsilon_2^{\dagger \alpha}(\pm)
\,{\cal M}_{\mu\alpha}
\,=\,
\frac{1}{m_1+m_2}\left(-P\cdot q\, A_0\mp 2\,m_1\,|{\bf p_2}|\, V \right),
\label{helS1c}\\
H_0 &=&
\epsilon^{\dagger \mu}(0)\,\epsilon_2^{\dagger \alpha}(0)
\,{\cal M}_{\mu\alpha}
\nonumber\\
&=&
\frac{1}{m_1+m_2}\frac{1}{2\,m_2\sqrt{q^2}}
\left(-P\cdot q\,(m_1^2-m_2^2-q^2)\, A_0
+4\,m_1^2\,|{\bf p_2}|^2\, A_+\right).\nonumber
\end{eqnarray}

\vspace{0.5cm}
\noindent
(c) \underline{$B_c\to (\bar cc)_{\,S=2}$ transition}: \\
\vspace{-0.1cm}

The nonvanishing helicity form factors can be calculated according to
\begin{equation}
H_m =\epsilon^{\dagger \mu}(m)\,{\cal M}_{\mu\alpha_1\alpha_2}\,
\epsilon_2^{\dagger\alpha_1\alpha_2}(m)
\quad {\rm for} \quad m=\pm,0
\end{equation}
and
\begin{equation}
H_t =\epsilon^{\dagger \mu}(t)\, {\cal M}_{\mu\alpha_1\alpha_2}\,
\epsilon_2^{\dagger\alpha_1\alpha_2}(0)
\end{equation}
Again the hadronic tensor is given by $H(m,n)= H_mH^\dagger_n$.

For the further evaluation one needs to specify the helicity components
$\epsilon_2(m)$ $(m=\pm 2,\pm 1,0)$ of the polarization vector
of the $\bar cc_{S=2}$.
They are given by
\begin{eqnarray}
\epsilon^{\mu\nu}_2(\pm 2) &=&\epsilon^\mu_2(\pm)\, \epsilon^\nu_2(\pm),
\nonumber\\
\epsilon^{\mu\nu}_2(\pm 1) &=&
\frac{1}{\sqrt{2}}\,\left(\epsilon^\mu_2(\pm)\,\epsilon^\nu_2(0)
                       +\epsilon^\mu_2(0)\,\epsilon^\nu_2(\pm)\right),
\label{pol_S2}\\
\epsilon^{\mu\nu}_2(0) &=&
\frac{1}{\sqrt{6}}\,\left(\epsilon^\mu_2(+)\,\epsilon^\nu_2(-)
                       +\epsilon^\mu_2(-)\,\epsilon^\nu_2(+)\right)
+\sqrt{\frac{2}{3}}\,\epsilon^\mu_2(0)\,\epsilon^\nu_2(0)\,,
\nonumber
\end{eqnarray}
where $\epsilon^\mu_2(r)$ are defined in Eq.~(\ref{pol_S1}).

The relation between the helicity form factors $H_m$ and
the invariant form factors Eq.~(\ref{ff2}) read
\begin{eqnarray*}
H_t &=& h(t,0)=\sqrt\frac{2}{3}\,
\frac{m_1^2\,  |{\bf p_2}|^2}{m_2^2\,\sqrt{q^2}}
\,\left\{T_1+( |{\bf p_2}|^2+E_2\, q_0+m_1\, q_0)\,T_2+q^2\, T_3\right\},
\\
H_\pm &=& h(\pm,\pm)=
\frac{m_1\,|{\bf p_2}|}{\sqrt{2}\,m_2}\,
\left(T_1\mp 2\, m_1|{\bf p_2}|\,T_4 \right),
\nonumber\\
H_0 &=& h(0,0)=
\sqrt\frac{1}{6}\,\frac{m_1\,|{\bf p_2}|}{m_2^2\,\sqrt{q^2}}
\left( (m_1^2-m_2^2-q^2)\, T_1 + 4\,m_1^2\,|{\bf p_2}|^2\,T_2 \right).
\end{eqnarray*}

\subsection{Lepton tensor}

The helicity components of the lepton tensors $L(m,n)$ are evaluated
in the ($l\bar\nu$)--c.m. system $\vec k_l+\vec k_\nu=0$.
One has
\begin{eqnarray}
q^\mu   &=& (\,\sqrt{q^2}\,,\,\,0\,,\,\,0\,,\,\,0\,)\,,
\nonumber\\
k^\mu_\nu &=&
(\,|{\bf k_l}|\,,\,\, |{\bf k_l}|\sin\theta\cos\chi\,,\,\,
|{\bf k_l}| \sin\theta\sin\chi\,,\,\,|{\bf k_l}| \cos\theta\,)\,,
\label{lepton_basis}\\
k^\mu_l &=& (\,E_l\,,\,\,-|{\bf k_l}|\sin\theta\cos\chi\,,\,\,
-|{\bf k_l}|\sin\theta\sin\chi\,,\,\,-|{\bf k_l}|\cos\theta\,)\,,
\nonumber
\end{eqnarray}
with
$$
E_l=\frac{q^2+\mu^2}{2\,\sqrt{q^2}}, \hspace{1cm}
|{\bf k_l}|=\frac{q^2-\mu^2}{2\,\sqrt{q^2}}.
$$
In the ($l\bar \nu$)--c.m. frame the longitudinal and time-component
polarization vectors are given by
\begin{eqnarray}
\label{lepton-pol}
\epsilon^\mu(t)  &=& \frac{q^\mu}{\sqrt{q^2}}=(1,0,0,0),
\nonumber\\
\epsilon^\mu(\pm)&=& \frac{1}{\sqrt{2}}\,(0,\mp,-i,0),
\label{hel_lep}\\
\epsilon^\mu(0) &=& (0,0,0,1).
\nonumber
\end{eqnarray}
Using Eqs.~(\ref{lepton-pol}) and (\ref{lepton-tensor}) it is not difficult
to evaluate the helicity representation $L(m,n)$ of the lepton
tensor.

In this paper we are not interested in the azimuthal $\chi$--distribution
of the lepton pair. We therefore integrate over the azimuthal angle
dependence of the
lepton tensor. Of course, our formalism is general enough to allow for the
inclusion of an azimuthal dependence if needed.
After azimuthal integration the differential $(q^2,\cos\theta)$
distribution reads
\begin{eqnarray}
\label{distr2}
\frac{d\Gamma}{dq^2d\cos\theta}
&=&\,
   \frac{3}{8}\,(1+\cos^2\theta)\cdot \frac{d\Gamma_U}{d q^2}
  +\frac{3}{4}\,\sin^2\theta\cdot     \frac{d\Gamma_L}{d q^2}
\mp\frac{3}{4}\cos\theta\cdot         \frac{d\Gamma_P}{d q^2}
\\
&& \nonumber \\
&+&
 \frac{3}{4}\,\sin^2\theta\cdot \frac{d\widetilde\Gamma_U}{d q^2}
+\frac{3}{2}\,\cos^2\theta\cdot \frac{d\widetilde\Gamma_L}{d q^2}
+\frac{1}{2}\,\frac{d\widetilde\Gamma_S}{d q^2}
+3\,\cos\theta\cdot \frac{d\widetilde\Gamma_{SL}}{d q^2},
\nonumber
\end{eqnarray}
where we take the polar angle $\theta$ to be the angle between
the $\vec p_2$ and the $\vec k_l$ in the lepton-neutrino c.m. system.
The upper and lower signs in front of the parity violating (p.v.)
contribution refer to the two cases $l^-\bar\nu$ and $l^+\nu$,
respectively.

The differential partial helicity rates
$d\Gamma_i/dq^2$ and $d\widetilde\Gamma_i/dq^2$
in Eq.~(\ref{distr2}) are defined by
\begin{eqnarray}
\label{partialrates}
\frac{d\Gamma_i}{dq^2} &=& \frac{G^2_F}{(2\pi)^3} \,
|V_{bc}|^2\,\frac{(q^2-\mu^2)^2\, |{\bf p_2}|}{12\,m_1^2\,q^2}\,H_i \, ,\\
&& \nonumber \\
\frac{d\widetilde\Gamma_i}{dq^2} &=&
\frac{\mu^2}{2\,q^2}\,\frac{d\Gamma_i}{dq^2} \, , \nonumber
\end{eqnarray}
where we have introduced a standard set of helicity structure
functions $H_i$ ($i=U,L,P,S,SL$) given by the following linear combinations
of the helicity components of the hadron tensor $H(m,n)=H_m H^\dagger_n$:
$$
\def\arraystretch{1.5}
\begin{array}{ll}
 H_{U}={\rm Re}\left(H_{+} H_{+}^{\dagger}\right)
            +{\rm Re}\left(H_{-} H_{-}^{\dagger}\right) &
\hspace*{1cm} {\rm {\bf U}npolarized-transverse} \\
% H_{T}={\rm Re}\left(H_{+} H_{-}^{\dagger}\right) &
%\hspace*{1cm} {\rm {\bf T}ransverse-interference} \\
 H_{L}={\rm Re}\left(H_{0} H_{0}^{\dagger}\right)  &
\hspace*{1cm}  {\rm{\bf L}ongitudinal} \\
H_{P}={\rm Re}\left(H_{+} H_{+}^{\dagger}\right)
       -{\rm Re}\left(H_{-} H_{-}^{\dagger}\right)  &
\hspace*{1cm} {\rm {\bf P}arity-odd} \\
 H_{S}=3\,{\rm Re}\left(H_{t} H_{t}^{\dagger}\right)  &
\hspace*{1cm}  {\rm{\bf S}calar} \\
 H_{SL}={\rm Re}\left(H_{t} H_{0}^{\dagger}\right)  &
 \hspace*{1cm} {\rm{\bf S}calar-{\bf L}ongitudinal \, interference}
\end{array}
$$
Note that the helicity amplitudes are real such that the complex
conjugation symbol can in fact be dropped.

It is evident that the ``tilde'' rates $\widetilde{\Gamma}$
in Eq.~(\ref{partialrates})
do not contribute
in the limit of vanishing lepton masses. In the present application
this means that they can be neglected for the $e$-- and
$\mu$--modes and only contribute to the $\tau$--modes.

Integrating over $\cos\theta$ one obtains the differential $q^2$ distribution
$$
\frac{d\Gamma}{dq^2} =
\frac{d\Gamma_U}{d q^2}+\frac{d\Gamma_L}{d q^2}
+\frac{d\widetilde\Gamma_U}{d q^2}+\frac{d\widetilde\Gamma_L}{d q^2}
+\frac{d\widetilde\Gamma_S}{d q^2}.
$$
Finally, integrating over $q^2$, one obtains the total rate
\[
\Gamma =
\Gamma_U+\Gamma_L+\widetilde\Gamma_U+\widetilde\Gamma_L+\widetilde\Gamma_S
\]
In Sec.~\ref{s:NumRes} we list our predictions for the integrated partial helicity
rates $\Gamma_i$ $(i=U, L, P)$ and $\widetilde\Gamma_i$ $i=(U, L, S, SL)$.

To save on notation in the following we shall sometimes use a
self-explanatory notation for the differential and integrated partial
helicity rates. For example, we write $U$ for either
the differential or the integrated helicity rates $d\Gamma_U/dq^2$ and
$\Gamma_U$, respectively, and $\widetilde{U}$ for $d\widetilde{\Gamma}_U/dq^2$ and
$\widetilde{\Gamma}_U$.

An interesting quantity
is the forward-backward asymmetry $A_{FB}$ of the lepton in the
($l\bar\nu$) cm system which is given by
\begin{equation}
\label{forward-backward}
A_{FB}=
\frac{3}{4} \frac{\pm P +4 \widetilde{SL}}{U+\widetilde{U}+L+\widetilde{L}+\widetilde{S}}.
\end{equation}
In Sec.~\ref{s:NumRes} we shall give our numerical predictions for the asymmetry $A_{FB}$
for the decay channels under study.

For the discovery channel $B_c\to J/\psi l \nu$ with the $J/\psi$ decaying
into muon pairs the transverse/longitudinal composition of the produced
$J/\psi$ is of interest. The transverse/longitudinal composition can be
determined by a measurement of the angular orientation of the back-to-back
muon pairs in the $J/\psi$ rest frame. The relevant angular distribution
reads
\begin{eqnarray}
\label{distr3}
\frac{d\Gamma}{dq^2d\cos\theta^*}
&=&\,
   \frac{3}{8}\,(1+\cos^2\theta^*) \left( \frac{d\Gamma_U}{d q^2} +
\frac{d\widetilde\Gamma_U}{d q^2} \right)                              \\
  &&+ \frac{3}{4}\,\sin^2\theta^* \left( \frac{d\Gamma_L}{d q^2}   +
\frac{d\widetilde\Gamma_L}{d q^2} + \frac{d\widetilde\Gamma_S}{d q^2} \right)
 \nonumber
\end{eqnarray}
where $\theta^*$ is the polar angle of the muon pair relative to the
original momentum direction of the $J/\psi$.
We have included lepton mass effects so that the angular decay distribution
in Eq.~(\ref{distr3}) can be also used for the $\tau$-mode in this decay.
The transverse and longitudinal contributions of the $J/\psi$ are
given by ($U + \widetilde{U}$) and ($L + \widetilde{L} +\widetilde{S} $),
respectively.

One can
define an asymmetry parameter $\alpha^*$ by rewriting Eq.~(\ref{distr2})
in terms of its $\cos^2\theta^*$ dependence, i.e.
$d\Gamma \propto 1+\alpha^*\cos^2\theta^*$. The asymmetry parameter
can be seen to be given by
\begin{equation}
\label{trans-long}
\alpha^*
=\frac{U+\widetilde{U}-2(L+ \widetilde{L}+\widetilde{S})}{U+\widetilde{U}+2(L + \widetilde{L}
+\widetilde{S})}\,.
\end{equation}
Our predictions for the asymmetry parameter $\alpha^*$ appear in Sec.~\ref{s:NumRes}.

We have only written out single angle decay distributions in this paper.
It is not difficult to write down joint angular decay distributions including
also azimuthal correlations in our formalism if necessary.

\section{Numerical results}
\label{s:NumRes}

%%%%%%%%%%%%%%%% 2nd insertion %%%%%%%%%%%%%%%

Let us discuss the model parameters and their
determination. Since we consider the decay of the $B_c$-meson
into charmonium states only, the adjustable parameters are
the constituent masses of charm and bottom quarks and
the size parameters of the $B_c$-meson and charmonium states.
The values of quark masses were determined in our previous
studies (see, for example, \cite{Ivanov:2003ge}) of the leptonic
and semileptonic decays of the low-lying pseudoscalar mesons
($\pi$, $K$, $D$, $D_s$, $B$, $B_s$ and $B_c$). The values
of the charm and bottom quarks were found to be $m_c=1.71$ GeV
and  $m_b=5.12$ GeV. The value of  $\Lambda_{bc}=1.96$ GeV
was determined from a fit to the world average of the
leptonic decay constant $f_{B_c}=360$ MeV.
The value of  $\Lambda_{J/\psi}=2.62$ GeV was found
from a fit to the experimental value of the radiative
decay constant $f_{J/\psi}=405$ MeV which enters in the
$J/\psi\to e^+ e^-$ decay width ($f_{J/\psi}^{\rm expt}=405\pm 17$ MeV).

In our calculation we are using free quark propagators
with an effective constituent quark mass (see, Eq.~\ref{local}).
This imposes a very simple yet important constraint
on the relations between the masses of the bound state
and their constituents. One has to assume that the meson mass $M_H$
is less than the sum of the masses of their constituents

\begin{equation}
\label{conf}
M_H < m_{q_1} + m_{q_2}
\end{equation}
in order to avoid the appearance of imaginary parts
in physical amplitudes, which are described by
the one-loop quark diagrams in our approach.
This is satisfied for the  low-lying pseudoscalar mesons
$\pi$, $K$, $D$, $D_s$, $B$, $B_s$, $B_c$ and
$\eta_c$ and also for the $J/\psi$ but is no longer
true for the excited charmonium states considered here.
We shall therefore employ identical masses for all
excited charmonium states $m_{cc}=m_{J/\psi}$=3.097 GeV
(except for the $\eta_c$) in our matrix element calculations
but use physical masses in the phase space calculation.
This is quite a reliable approximation because  the
hyperfine splitting between the excited charmonium states
and  $J/\psi$ is not large. For example,
the maximum relative  error is
$(m_{\psi(3836)}-m_{J/\psi})/m_{J/\psi}=0.24$.

The size parameters of the excited charmonium states
should be determined from a fit to the
available experimental data for the two-photon
and the radiative decays as was done for the $J/\psi$-meson.
However, the calculation of the matrix elements
involving two photons will be very time
consuming because one has to introduce the
electromagnetic field into the nonlocal Lagrangians
in Eqs.~\ref{eq:s=0}-\ref{eq:s=2}. This is done  by using
the path exponential (see, our recent papers
\cite{Ivanov:2003ge} and \cite{Faessler:2003yf}).
The gauging of the nonlocal Lagrangian with spin 2
has not yet been done and is a project all of its own.
For the time being, we are  calculating  the widths of
the semileptonic decays $B_c\to (\bar cc) +l\nu$
by assuming an identical size parameters
for all charmonium states
$\Lambda_{cc}=\Lambda_{J/\psi}=2.62$ GeV.

%%%%%%%%%%%%%%%%%%%%%%%%%%%%%%%

In order to get a quantitative idea about the invariant form factors
we list their $q^2_{\rm min}=0$ and $q^2_{\rm max}=(m_1-m_2)^2$ values in
Table~\ref{t:formfactors}.
\begin{table}[ht]
\begin{center}
\def\arraystretch{1.5}
\caption {\label{t:formfactors} Predictions for the form factors
of the $B_c\to (\bar cc)$ tansitions. }
\begin{tabular}{|l|c|c|c|}
\hline
              & $q^2$           & $F_+$ & $F_-$ \\
\hline\hline
 $\eta_c$     & 0               & 0.61  & -0.32    \\
              & $q^2_{\rm max}$ & 1.14  & -0.61    \\
\hline
  $\chi_{c0}$ & 0               &  0.40 & -1.00   \\
              & $q^2_{\rm max}$ &  0.65 & -1.63   \\
\hline\hline
\end{tabular}

\vspace{0.3truecm}

\begin{tabular}{|l|c|c|c||c|c|}
\hline
           & $q^2$ & $A_0$ & $A_+$ & $A_-$  & $V$ \\
\hline\hline
 $J/\psi$    & 0   &  1.64  &  0.54 & -0.95  &  0.83 \\
 & $q^2_{\rm max}$ &  2.50  &  0.97 & -1.76  &  1.53 \\
\hline
 $\chi_{c1}$   & 0 & -0.064 & -0.39 &  1.52  & -1.18  \\
 & $q^2_{\rm max}$ &  0.46  & -0.50 &  2.36  & -1.81 \\
\hline
 $h_c$     & 0     & 0.44  & -1.08 &  0.52  &  0.25 \\
 & $q^2_{\rm max}$ & 0.54  & -1.80 &  0.89  &  0.365 \\
\hline\hline
\end{tabular}

\vspace{0.3truecm}

\begin{tabular}{|l|c|c|c|c|c|}
\hline
 & $q^2$ & $T_1$ &
 $T_2$,\,GeV$^{-2}$ & $T_3$,\,GeV$^{-2}$ & $T_4$,\,GeV$^{-2}$ \\
\hline\hline
 $\chi_{c2}$ & 0    & 1.22  & -0.011  & 0.025    & -0.021  \\
 & $q^2_{\rm max}$  & 1.69  & -0.018  & 0.040    & -0.033\\
\hline
 $\psi(3836)$& 0    & 0.052 &  0.0071 &  -0.036  &  0.026 \\
 & $q^2_{\rm max}$  & 0.35  &  0.0090 &  -0.052  &  0.038 \\
\hline\hline
\end{tabular}
%\vspace{-0.1truecm}
\end{center}
\end{table}

%%%%%%%%%  1-st insertion

We put our values of the decay rates in
Table~\ref{t:widths} together with those predicted
in other papers.  A number of calculations are devoted to
the $B_c\to \eta_c l\nu$ and $B_c\to J/\psi l\nu$
decays. All of them predict values
at the same order of magnitude. A  study
of the semileptonic decays of the $B_c$-meson
into excited charmonium states was done in
\cite{Chang:1992pt,Chang:2001pm} within an
approach which is quite different from our
relativistic quark model.
Concerning the electron-modes here is quite good agreement
with \cite{Chang:1992pt,Chang:2001pm} in the case of
$B_c\to J/\psi e\nu$, $\eta_c e\nu$, $\chi_{c2} e\nu$.
Our rates are a factor of 1.5 (1.8) larger
for $B_c\to \chi_{c0} e\nu$, $h_c e\nu$ decays
and our rate is a factor 1.6 smaller for
$B_c\to\chi_{c1} e\nu$ decay.
Concerning the $\tau$-modes here is quite good agreement
with \cite{Chang:1992pt,Chang:2001pm} in the case of
$B_c\to\chi_{c0}\tau\nu$, $h_c\tau\nu$ decays but our rates
are almost a factor of two smaller
for the other modes $B_c\to\chi_{c1}\tau\nu$, $\chi_{c2}\tau\nu$.
%
%%%%%

The partial rate for $B_c\to J/\psi+l+\nu$ is the largest. The partial
rates into the P-wave charmonium states are all of the same order of
magnitude and are predicted to occur at $\sim 10\%$ of the most
prominent decay $B_c\to J/\psi+l+\nu$. The decays of the $B_c$ into D-wave
charmonium state are suppressed. The $\tau$--modes are generally down by
a factor of $\sim 10$ compared to the $e$--modes except for the
transitions $B_c\to\eta_c$,
$B_c\to J/\psi$ and $B_c\to \psi(3836)$ where the $\tau$--modes
are smaller only by a factor of $\sim 3-4$.

In Table~\ref{t:hel} we list our results for the integrated partial
helicity rates $\Gamma_i$
$(i=U,L,P,\widetilde{U}, \widetilde{L}, \widetilde{S},\widetilde{SL})$.
They are needed
for the calculation of the forward-backward asymmetry parameter $A_{FB}$
and, in the case of the decay $B_c\to J/\psi+l+\nu$, for the calculation
of the asymmetry
parameter $\alpha^*$ determining the transverse/longitudinal
composition of the $J/\psi$ in the decay. The partial ``tilde'' rates
$\widetilde{\Gamma}_i$ are quite tiny for the $e$--mode as expected from
Eq.~(\ref{partialrates}) but are not negligible for
the $\tau$--modes. This shows up in the calculated values for $A_{FB}$ in
Table~\ref{t:AFB}. For the decays into spin $0$ states $A_{FB}$ is
proportional to $\widetilde{SL}$ and thus tiny for the $e$--mode but
nonnegligible for the $\tau$--modes. For the decay into the other spin
states one has $A_{FB}(e^-)= - A_{FB}(e^+)$ but
$A_{FB}(\tau^-)\ne - A_{FB}(\tau^+)$ as can easily be appreciated by looking
at Eq.~(\ref{forward-backward}). The forward-backward asymmetry can amount
up to 40 \%. The transverse and the longitudinal
pieces of the $J/\psi$ in the decay $B_c\to J/\psi+l+\nu$ are almost equal
for both the $e$-- and the $\tau$--modes (see Table~\ref{t:hel}). According
to Eq.~(\ref{trans-long}) this implies that the asymmetry parameter
$\alpha^*$ should be close to $- 33 \%$ as is indeed the case as the entries
in Table~\ref{t:AFB} show. For the other two modes involving spin 1
charmonium states the transverse/longitudinal population is quite different.
For the transition $B_c\to \chi_{c1}$ the transverse mode dominates by a
factor of $\sim 3$ for both the $e$-- and $\tau$--modes whereas for the
transition $B_c\to h_c$ the longitudinal mode dominates by a factor of
$\sim 13$ and $\sim 7$ for the $e$-- and $\tau$--modes, respectively.

Taking the central value of the CDF lifetime measurement $\tau(B_c)=0.46
\cdot 10^{-12}$ s \cite{CDF} and our predictions for the rates into
the different charmonium states one finds branching fractions of
$\sim 2 \%$ and $\sim 0.7 \%$ for the decays into the two $S$--wave
charmonium states $J/\psi$ and $\eta_c$, respectively, and branching fractions of
$\sim 0.2 \%$ for the deacys into the $P$--wave charmonium states.
Considering the fact that there will be a yield of up to $10^{10}$ $B_c$
mesons per year at the Tevatron and LHC the semileptonic decays of the
$B_c$ mesons into charmonium states studied in this paper offers a
fascinating area of future research.

\begin{table}[ht]
\begin{center}
\caption{\label{t:widths}
Semileptonic decay rates in units of $10^{-15}$ GeV. We use $|V_{cb}|= 0.04$.}
\def\arraystretch{1.5}
\begin{tabular}{|l|c|c|c|c|c|c|c|}
\hline\hline
 & This model & \cite{Chang:1992pt,Chang:2001pm} &
\cite{AMV} &\cite{AKNT} & \cite{KLO} & \cite{Colangelo} &
\cite{Ebert:2003cn} \\
\hline\hline
$B_c\to\eta_c\, e\,\nu$         & 10.7  & 14.2 & 11.1 & 8.6 & 11$\pm$1 &
2.1 (6.9) & 5.9 \\
\hline
$B_c\to\eta_c\,\tau\,\nu$       & 3.52  & & & 3.3$\pm$0.9 & & & \\
\hline\hline
$B_c\to J/\psi\, e\,\nu$        & 28.2  & 34.4 & 30.2 & 17.2 & 28$\pm$5 &
21.6 (48.3) & 17.7 \\
\hline
$B_c\to J/\psi \,\tau\,\nu$     & 7.82  & & & 7$\pm$2     & & &  \\
\hline\hline
$B_c\to \chi_{c0}\, e\,\nu$     & 2.52  & 1.686 & & & & &   \\
\hline
$B_c\to\chi_{c0} \,\tau\,\nu$   & 0.26  & 0.249  & & & & &  \\
\hline\hline
$B_c\to \chi_{c1} \, e\,\nu$    & 1.40  & 2.206 & & & & & \\
\hline
$B_c\to \chi_{c1} \,\tau\,\nu$  & 0.17  & 0.346 & & & & &\\
\hline\hline
$B_c\to h_c\, e\,\nu $          & 4.42  & 2.509 & & & & &\\
\hline
$B_c\to h_c\, \tau\,\nu$        & 0.38  & 0.356 & & & & &\\
\hline\hline
$B_c\to \chi_{c2}\, e\,\nu $    & 2.92  & 2.732 & & & & &\\
\hline
$B_c\to \chi_{c2}\, \tau\,\nu$  & 0.20  & 0.422 & & & & &\\
\hline\hline
$B_c\to \psi(3836)\, e\,\nu $   & 0.13  & & & & & &\\
\hline
$B_c\to \psi(3836)\, \tau\,\nu$ & 0.0031 &  & & & & &\\
\hline\hline
\end{tabular}
\end{center}
\end{table}

\begin{table}[ht]
\begin{center}
\def\arraystretch{1.5}
\caption{\label{t:hel}
Partial helicity rates in units of $10^{-15}$ GeV.}
\begin{tabular}{|l|c|c|c|c|c|c|c|}
\hline\hline
 Mode & $U$ & $\widetilde U$ & $L$ & $\widetilde L$ &$P$ & $\widetilde S$&$\widetilde {SL}$ \\
\hline\hline
$B_c\to\eta_c\, e\,\nu$         & 0 & 0 & 10.7 & $0.34\,10^{-5}$ &
                                  0 & $0.11\,10^{-4}$ & $0.34\,10^{-5}$  \\
\hline
$B_c\to\eta_c\,\tau\,\nu$       & 0  & 0 & 1.18 & 0.27 & 0&  2.06 & 0.42 \\
\hline\hline
$B_c\to J/\psi\, e\,\nu$        & 14.0 & $3.8\,10^{-7}$ & 14.2 &
                                  $0.31\,10^{-5}$ & 8.00 & $0.88\,10^{-5}$
                                & $0.30\,10^{-5}$\\
\hline
$B_c\to J/\psi \,\tau\,\nu$     & 3.59 & 0.73 & 2.35 & 0.50 & 1.74 & 0.63
                                & 0.31   \\
\hline\hline
$B_c\to \chi_{c0}\, e\,\nu$     & 0 & 0 & 2.52 & $0.11\,10^{-5}$ & 0
                                & $0.32\,10^{-5}$ & $0.11\,10^{-5}$    \\
\hline
$B_c\to\chi_{c0} \,\tau\,\nu$   & 0 & 0 & 0.13 & 0.037 & 0 & 0.089 & 0.033  \\
\hline\hline
$B_c\to \chi_{c1} \, e\,\nu$    & 1.08  & $0.51\,10^{-7}$ & 0.33 &
                                 $0.11\,10^{-6}$ & -0.35 & $0.32\,10^{-6}$
                                & $0.11\,10^{-6}$ \\
\hline
$B_c\to \chi_{c1} \,\tau\,\nu$  & 0.11 & 0.029 & 0.024 & 0.0064 &
                                  -0.066 & $0.42\,10^{-2}$&$0.29\,10^{-2}$ \\
\hline\hline
$B_c\to h_c\, e\,\nu $          & 0.33 & $0.13\,10^{-7}$ & 4.10 &
                                 $0.24\,10^{-5}$ & 0.21 & $0.72\,10^{-5}$
                                & $0.24\,10^{-5}$\\
\hline
$B_c\to h_c\, \tau\,\nu$        & 0.045 & 0.012 & 0.13 & 0.037 & 0.023 &
                                   0.15 & 0.044 \\
\hline\hline
$B_c\to \chi_{c2}\, e\,\nu $    & 0.98 & $0.53\,10^{-7}$ & 1.93 &
                                   $0.10\,10^{-5}$ & 0.62 &
                                   $0.30\,10^{-5}$ & $0.99\,10^{-6}$\\
\hline
$B_c\to \chi_{c2}\, \tau\,\nu$  & 0.073 & 0.021 & 0.066 & 0.019 & 0.036
                                &  0.025 & 0.012 \\
\hline\hline
$B_c\to \psi(3836)\, e\,\nu $   & 0.075 & $0.06\,10^{-7}$ & 0.052
                                & $0.36\,10^{-7}$ & -0.036 &$0.10\,10^{-6}$
                                & $0.35\,10^{-7}$ \\
\hline
$B_c\to \psi(3836)\, \tau\,\nu$ & $0.16\,10^{-2}$ & $0.53\,10^{-3}$ &
                               $0.66\,10^{-3}$ & $0.22\,10^{-3}$ &
                               $-0.13\,10^{-2}$ & $0.16\,10^{-3}$
                                & $0.11\,10^{-3}$ \\
\hline\hline
\end{tabular}
\end{center}
\end{table}

\begin{table}[ht]
\begin{center}
\def\arraystretch{1.5}
\caption{\label{t:AFB}
Forward-backward asymmetry $A_{FB}$ and the asymmetry
parameter $\alpha^\ast$.}
\begin{tabular}{|l|c|c|c|}
\hline\hline
 Mode & $A_{FB}(l^-)$ & $A_{FB}(l^+)$ & $\alpha^\ast$ \\
\hline\hline
$B_c\to\eta_c\, e\,\nu$         & $9.64\,10^{-6}$ & $9.64\,10^{-6}$ & - \\
\hline
$B_c\to\eta_c\,\tau\,\nu$       &  0.36           &  0.36           & - \\
\hline\hline
$B_c\to J/\psi\, e\,\nu$        &  0.21           & -0.21           & -0.34 \\
\hline
$B_c\to J/\psi \,\tau\,\nu$     &  0.29           & -0.05           & -0.24 \\
\hline\hline
$B_c\to \chi_{c0}\, e\,\nu$     & $1.29\,10^{-6}$ & $1.29\,10^{-6}$ & - \\
\hline
$B_c\to\chi_{c0} \,\tau\,\nu$   &  0.38           &  0.38           & - \\
\hline\hline
$B_c\to \chi_{c1} \, e\,\nu$    & -0.19           & 0.19            & - \\
\hline
$B_c\to \chi_{c1} \,\tau\,\nu$  & -0.24           & 0.34            & -  \\
\hline\hline
$B_c\to h_c\, e\,\nu $          & 0.036           & -0.036          & - \\
\hline
$B_c\to h_c\, \tau\,\nu$        & 0.39            & 0.30            & - \\
\hline\hline
$B_c\to \chi_{c2}\, e\,\nu $    & 0.16            & -0.16           & - \\
\hline
$B_c\to \chi_{c2}\, \tau\,\nu$  & 0.32            & 0.05            & - \\
\hline\hline
$B_c\to \psi(3836)\, e\,\nu $   & -0.21           & 0.21            & -0.17 \\
\hline
$B_c\to \psi(3836)\, \tau\,\nu$ & -0.21           & 0.41            & 0.006 \\
\hline\hline
\end{tabular}
\end{center}
\end{table}

\section*{Acknowledgments}
\noindent
M.A.I. appreciates the partial support by the DFG
(Germany) under the grant 436RUS17/26/04, the Heisenberg-Landau Program
and the Russian Fund of Basic Research (Grant No.04-02-17370).

\appendix
\section*{Appendix: Convention for Dirac $\gamma$-matrices and
the antisymmetric tensor in  Minkowski space}

We use the conventions of Bjorken-Drell. Thus we define the metric tensor
and the totally antisymmetric
$\varepsilon$-tensor in Minkowski space by
$g^{\mu\nu}=g_{\mu\nu}={\rm diag}(+,-,-,-,)$ and $\varepsilon_{0123}=
-\varepsilon^{0123}= 1$.
For the partial and full contractions of a pair of $\varepsilon$-tensors
one finds
\begin{eqnarray*}
\varepsilon_{\mu_1\mu_2\mu_3\mu_4}\varepsilon^{\nu_1\nu_2\nu_3\mu_4}&=&
-g_{\mu_1}^{\nu_1}g_{\mu_2}^{\nu_2}g_{\mu_3}^{\nu_3}
-g_{\mu_1}^{\nu_2}g_{\mu_2}^{\nu_3}g_{\mu_3}^{\nu_1}
-g_{\mu_1}^{\nu_3}g_{\mu_2}^{\nu_1}g_{\mu_3}^{\nu_2}
\\
&&
 +g_{\mu_1}^{\nu_1}g_{\mu_2}^{\nu_3}g_{\mu_3}^{\nu_2}
+g_{\mu_1}^{\nu_2}g_{\mu_2}^{\nu_1}g_{\mu_3}^{\nu_3}
+g_{\mu_1}^{\nu_3}g_{\mu_2}^{\nu_2}g_{\mu_3}^{\nu_1}
\\
\varepsilon_{\mu_1\mu_2\mu_3\mu_4}\varepsilon^{\nu_1\nu_2\mu_3\mu_4}&=&
 -\,2\,(g_{\mu_1}^{\nu_1}g_{\mu_2}^{\nu_2}
      - g_{\mu_1}^{\nu_2}g_{\mu_2}^{\nu_1})
\\
\varepsilon_{\mu_1\mu_2\mu_3\mu_4}\varepsilon^{\nu_1\mu_2\mu_3\mu_4} &=&
- 6\,g_{\mu_1}^{\nu_1}
\\
\varepsilon_{\mu_1\mu_2\mu_3\mu_4}\varepsilon^{\mu_1\mu_2\mu_3\mu_4} &=& -24
\end{eqnarray*}

We employ the following definition of the $\gamma^5$-matrix
\begin{eqnarray*}
&&
\gamma^5 = \gamma_5= i\, \gamma^0\gamma^1\gamma^2\gamma^3
=\frac{i}{24}\varepsilon_{\mu_1\mu_2\mu_3\mu_4}\,
 \gamma^{\mu_1}\gamma^{\mu_2}\gamma^{\mu_3}\gamma^{\mu_4}=
\left(
\begin{array}{lr}
 0 & I \\
 I & 0 \\
\end{array}
\right),
\\
&&\\
&&
{\rm Tr}
\left(\gamma_5\gamma^{\mu_1}\gamma^{\mu_2}\gamma^{\mu_3}\gamma^{\mu_4}\right)
= 4\,i\,\varepsilon^{\mu_1\mu_2\mu_3\mu_4}.
\end{eqnarray*}

The leptons with negative charge ($l=e^-,\,\mu^-,\,\tau^-$) are referred
to as ``leptons'' whereas the positively charged leptons
$\bar l=e^+,\,\mu^+,\,\tau^+$ are referred to as
``antileptons''.

\end{document}